\documentclass[fleqn,twoside]{article}
\usepackage{espcrc2}

\usepackage{graphicx}
\usepackage[figuresright]{rotating}

\newcommand{\AmS}{{\protect\the\textfont2
  A\kern-.1667em\lower.5ex\hbox{M}\kern-.125emS}}

\title{The $\Upsilon$ Spectrum from Lattice QCD with 2+1 Flavors of Dynamical Quarks}

\author{A. Gray\address[UG]{Dept of Physics and Astronomy, 
        University of Glasgow, Glasgow G12 8QQ, UK},
        C. Davies\addressmark[UG],
        M. Alford\addressmark[UG],
	E. Follana\addressmark[UG],
	J. Hein\address[EU]{School of Physics, 
        University of Edinburgh, Edinburgh EH9 3JZ, UK},
	G. P. Lepage\address[CU]{Newman Laboratory of Nuclear Studies, Cornell University, Ithaca NY 14853, USA},
	Q. Mason\addressmark[CU],
	M. Nobes\address[SFU]{Department of Physics, Simon Fraser University,  Burnaby, British Columbia V5A 1S6, CANADA},
	J. Shigemitsu\address[OSU]{Department of Physics, the Ohio State University, OH 43210, USA},
	H. Trottier\addressmark[SFU],
	M. Wingate\addressmark[OSU],
	HPQCD and UKQCD collaborations.}
       
\begin{document}

\begin{abstract}
We describe the bottomonium spectrum obtained on the MILC configurations which incorporate 2+1 flavors of dynamical quarks. We compare to quenched and 2 flavor results also on MILC configurations. We show that the lattice spacing determination using different quantities shows clear signs of convergence with 2+1 flavors and give results for the leptonic width and hyperfine splitting, in the form of the ratio of the 1st excited state of the $\Upsilon$ to that of the ground state.
\vspace{1pc}
\end{abstract}

\maketitle

\section{INTRODUCTION}

Measurement of the $\Upsilon$ spectrum gives an excellent test for studying the dynamical content of the configurations being used. NRQCD can be used to give precise results which indicate how the light dynamical quarks feed in to the heavy meson properties. The program at CLEO-c will enable lattice results on $\Upsilon$ spectroscopy to be experimentally verified \cite{Gal02}.

Here we present new results using dynamical configurations from the MILC collaboration \cite{MILC01}. We have used the coarser set of available ensembles. These were generated using the improved staggered Asqtad action and have the advantage over previous configurations of the incorporation of 2+1 flavors of dynamical quarks. These are therefore the most realistic simulations of the physical vacuum to date. Several 2+1 flavor ensembles are available  with one dynamical quark being held at approximately the value of the strange quark mass $m_s$ and two degenerate flavors varying from $m_s$  to $m_s/5$. Also available are a quenched ensemble and a two flavor ensemble with degenerate dynamical quark masses of $m_s/2.5$. We have used both the most chiral 2+1 flavor ensemble and the degenerate 3 flavor ensemble plus the quenched and 2 flavor ensembles to best demonstrate the effects of unquenching on the bottomonium spectrum. 

We use the standard ${\cal{O}}(v^4)$ NRQCD Lagrangian \cite{Dav94}
\begin{eqnarray}
{\cal L} & = & \psi^{\dagger}(D_t+H_0)\psi+\psi^{\dagger}\delta H\psi; \\
H_0 & = & -\frac{\Delta^{(2)}}{2M}\\
\delta H & = & -\frac{({\bf{\Delta}}^{(2)})^2}{8M^3}+\frac{ig}{8M^2}({\bf{\Delta}}^{(\pm)}\cdot{\bf{E}}-{\bf{E}}\cdot{\bf{\Delta}}^{(\pm)})\nonumber \\
& - & \frac{g}{8M^2}{\bf{\sigma}}\cdot({\bf{\Delta}}^{(\pm)}\times{\bf{E}}-{\bf{E}}\times{\bf{\Delta}}^{(\pm)}) \nonumber \\ & - & \frac{g}{2M}{\bf{\sigma}}\cdot{\bf{B}} 
 +  \frac{a^2\Delta^{(4)}}{24M}-\frac{a(\Delta^{(2)})^2}{16nM^2}  
\end{eqnarray}
but with corrections to the $E$ and $B$ fields for ${\cal{O}}(a^2)$ errors. All terms are tadpole improved using an estimate of the Landau gauge link.
\begin{figure}[t]
\includegraphics[scale=0.375]{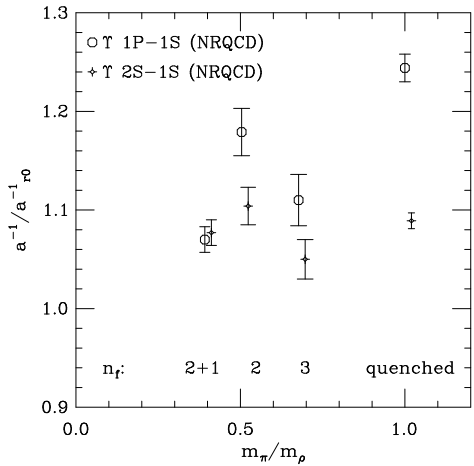}
\caption{The ratio of the inverse lattice spacing obtained from spectroscopy to that obtained from $r_0$. Open circles and fancy diamonds represent scales determined from the 1P-1S $\Upsilon$ and 2S-1S $\Upsilon$ splittings respectively.}
\label{fig:ainv}
\end{figure}
\section{FITTING}
For all ensembles, 210 configurations were used. For all but the degenerate 3 flavor ensemble, 4 different quark (antiquark) origins were used on each of  4 different timeslices. For the degenerate 3 flavor ensemble, 4 different origins were used  on each of 2 different timeslices.
In fitting the meson correlators, the Bayes technique \cite{Lep02} was used. 
$3\times3$ matrix fits were done using local, ground and excited state smearings for the S states and $2\times2$ matrix fits were done using ground and excited state smearings for P states. The Bayes parameters used were quite relaxed with  $ln(E_0)=-1.1(7)$, each of the amplitudes = 0.1(7), and splittings between excited states $\sim 500\mbox{MeV}\pm\mbox{ factor 2}$.
\section{RESULTS FOR THE LATTICE SPACING}
In Figure \ref{fig:ainv} results are plotted for the ratio of $a^{-1}$ obtained from both the 1P-1S and 2S-1S splittings in the $\Upsilon$ system to that obtained from using $r_0$, assuming that $r_0=0.5\mbox{fm}$. The results from the dynamical configurations are plotted against the ratio of the pseudoscalar to vector light-light meson mass where the valence quark mass is equal to the $u/d$ sea quark mass. We will also be able to extract a reliable scale from the quantity $2m_{B_s}-m_{\Upsilon}$ once $m_s$ is well determined \cite{WIN02}. Dependence on the quantity used to obtain $a^{-1}$ with quenching gives ambiguities in calculating other physical observables. The plot clearly illustrates these ambiguities disappearing as $m_{\pi}/m_{\rho}$ tends to its physical value. For the most chiral 2+1 flavor dynamical ensemble the scales from $\Upsilon$ have converged.
The $\Upsilon$ results imply that $r_0<0.5\mbox{fm}$. These scales can be used to determine a value for $\alpha_s$ \cite{Dav02}.

\section{RESULTS FOR THE LEPTONIC WIDTH AND HYPERFINE SPLITTING}
\begin{figure}[t]
\includegraphics[scale=0.375]{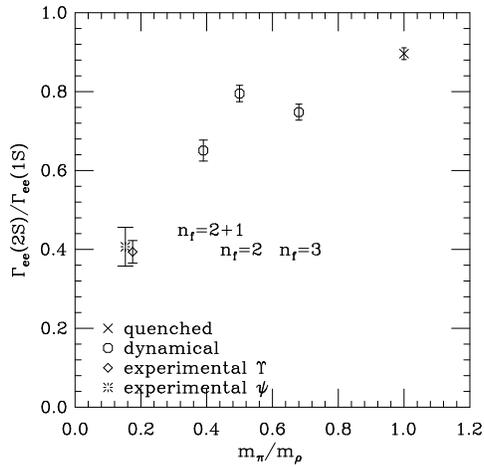}
\caption{The ratio of the leptonic width for the 1st radially excited state of  $\Upsilon$ to that of the ground state. The open circles represent results for dynamical configurations while the cross represents the quenched result. The diamond and burst represent the current experimental results for $\Upsilon$ and $\psi$ respectively. }
\label{fig:leprat}
\vspace*{-0.6cm}
\end{figure}

Figure \ref{fig:leprat} shows the ratio of the leptonic width, i.e. the width of decay to $e^+e^-$, of the 1st radially excited $\Upsilon$ to that of the ground state  $\Upsilon$. The experimental points for $\psi$ and $\Upsilon$ are plotted to demonstrate that this ratio is to a large extent independent of the heavy quark mass. At leading order $\Gamma_{ee}(nS)\propto|\psi_n(0)|^2$, where the `wavefunction at the origin'  $\psi_n(0) = \langle n|l\rangle$, the overlap of the local vector current with a given state. Therefore we can calculate this ratio using appropriate amplitudes from our $3\times3$ fits. The $Z$ factor for this current, which must be calculated in perturbation theory, cancels in the ratio. It can be seen that although even the most chiral result is still some way from experiment, the ratio is clearly falling with dynamical quark mass. Next, the currents must be improved by adding current correction operators as has been done for $f_B$ \cite{Morn98}.

\begin{figure}[t]
\includegraphics[scale=0.375]{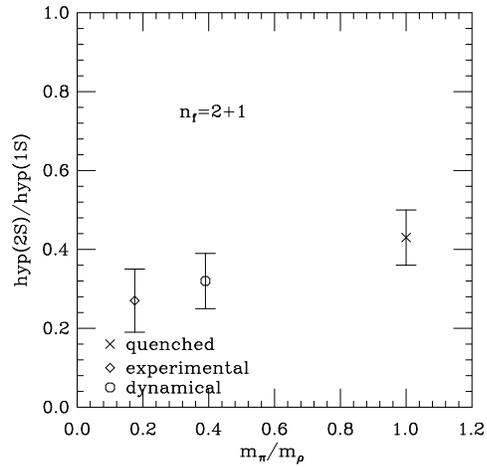}
\caption{The ratio of the hyperfine splitting for the 1st excited state of the $\Upsilon$ to that of the ground state. The open circle represents the results for the most chiral dynamical configuration while the cross represents the quenched result. The diamond represents the current experimental result for the $\psi$ system. }
\label{fig:hyprat}
\vspace*{-0.60cm}
\end{figure}

The hyperfine splitting between the $\Upsilon$ and $\eta_b$ is a spin dependent quantity which is proportional to the square of the coefficient of the $\sigma.B$ term in the action and also, in leading order in a potential model, to $|\psi_n(0)|^2$.
In Figure \ref{fig:hyprat} we show the ratio of hyperfine splitting of the 1st excited to that of the ground $\Upsilon$ state. The uncertainty from the coefficient of the $\sigma.B$ term, and, similarly to the leptonic width, a lot of the heavy quark mass dependence cancels in the ratio.
The $\eta_b$ and $\eta_b'$ are experimentally  unseen as yet so the experimental result given is for the $\psi$ system from the BELLE collaboration \cite{Belle}.
The plot shows a tendency towards experiment with unquenching.


\section{CONCLUSION}


The dependence of the scale on the quantity used to obtain it has been a long standing problem of the quenched approximation. Full convergence of $a^{-1}$ obtained from different splittings in the $\Upsilon$ spectrum is now clearly seen on the most chiral 2+1 flavor MILC ensemble. We conclude that this ensemble has good dynamical content. Ratios of leptonic widths and hyperfine splitting have been measured and are seen to tend toward experiment wih unquenching and decreasing dynamical quark mass.

\section{ACKNOWLEDGEMENTS}
We are grateful to MILC for use of their configurations. This work was supported by NSF, DoE, PPARC, the EU, and NSERC Canada.

\end{document}